\documentclass[aps,pre,twocolumn,superscriptaddress,showpacs]{revtex4}
\usepackage{graphicx}

\begin{document}

\title{Fast ignition of inertial fusion targets by laser-driven carbon beams}

\author{J.J.~Honrubia}
\email{javier.honrubia@upm.es}
\affiliation{ETSI Aeron\'auticos, Universidad Polit\'ecnica de Madrid, Madrid-28040, Spain}

\author{J.C.~Fern\'andez}
\affiliation{Los Alamos National Laboratory, Los Alamos, NM-87544, USA}

\author{M.~Temporal}
\affiliation{ETSI Aeron\'auticos, Universidad Polit\'ecnica de Madrid, Madrid-28040, Spain}

\author{B.M.~Hegelich}
\affiliation{Los Alamos National Laboratory, Los Alamos, NM-87544, USA}
\affiliation{Fakult\"at f\"ur Physik, LMU M\"unchen, D-85748 Garching, Germany}

\author{J.~Meyer-ter-Vehn}
\affiliation{Max-Planck-Institut f\"ur Quantenoptik, D-85748 Garching, Germany}

\date{\today}

\begin{abstract}
Two-dimensional simulations of ion beam driven fast ignition
are presented. Ignition energies of protons with Maxwellian
spectrum and carbon ions with quasimonoenergetic and Maxwellian
energy distributions are evaluated. The effect of the coronal
plasma surrounding the compressed deuterium-tritium is studied
for three different fuel density distributions. It is found that
quasimonoenergetic ions have better coupling with the compressed
deuterium-tritium and substantially lower ignition energies.
Comparison of quasimonoenergetic carbon ions and relativistic
electrons as ignitor beams shows similar laser energy requirements,
provided that a laser to quasimonoenergetic carbon ion conversion
efficiency around 10\% can be achieved.
\end{abstract}


\maketitle

\section{Introduction}

Fast ignition by laser-driven ion beams has been proposed
\cite{Roth, Roth2} as an alternative to the standard
scheme of fast ignition by relativistic electrons \cite{Tabak}.
It offers the advantages of the classical interaction of
ion beams with compressed fuels, their much more localized
energy deposition, and the stiffer ion transport with the
possibility of beam focusing. Proton fast ignition (PFI)
is a promising option because of the high laser-to-proton
conversion efficiencies ($\approx$ 6 - 12\%) found in
experiments for Maxwellian energy distributions \cite{Robson,Snavely}.
Novel target geometries such as flat-top cone targets seem to be
appropriate to increase those efficiencies \cite{Flippo}.
Monoenergetic protons may have a better coupling with the compressed
core, but the relatively low conversion efficiencies ($\approx$ 1\% 
\cite{Fuchs}) found so far hamper their application to fast ignition.

Studies on PFI have shown that ignition energies can be
reduced substantially by using a sequence of two beams \cite{Temporal2}.
Numerical simulations show that fuel targets compressed to
$\rho$ = 500 g/cm$^3$ can be ignited by two Maxwellian proton
beams with temperature T$_p$ = 4 MeV and a total energy of 8 kJ
\cite{Temporal3}. Assuming a conversion efficiency of 10\%, the
required laser energy is about 80 kJ, that is of the same
order than the energies envisioned for future fast ignition facilities
such as HiPER (High Power laser Energy Research) \cite{Dunne}.
However, the PFI scheme requires
to place the proton source near the compressed fuel inside a
re-entrant cone to shield the source from the plasma coming from
the imploding shell. Thus, the capsule implosion and fuel
compression will be relatively complex due to the flow
perturbations induced by the cone.
A complete simulation of an indirectly driven target ignited
by laser-driven protons can be found in \cite{Ramis}.

Recently, particle-in-cell (PIC) simulations of the interaction
of short pulses of circularly polarized laser light with thin
foils at intensities of 10$^{22}$ W/cm$^2$ have shown that
it is possible to accelerate ions up to hundreds of MeV with
quasimonoenergetic energy distributions and small divergence
angles \cite{Macchi, Klimo, Robinson, Rykovanov, Naumova, Qiao,
Chen}. In this scheme, that we refer to as RPA (Radiation
Pressure Acceleration),
the whole foil is accelerated by radiation pressure
with laser-to-ion conversion efficiencies around 10\% \cite{Klimo}.
The so-called laser break-out afterburner (BOA) scheme has
been also proposed to generate almost perfectly collimated
quasimonoenergetic ions with conversion efficiencies of a few
percent \cite{Yin} at laser intensities about 10$^{21}$ W/cm$^2$.
These new schemes are suitable to accelerate
ions heavier than protons to the energies required for fast
ignition applications. Fern\'andez et al. \cite{Fernandez, Fernandez2}
suggested the use of quasimonoenergetic carbon ions with kinetic
energies of a few hundreds of MeV to ignite pre-compressed
fusion targets \cite{Albright}. Use of heavier ions has been
studied in Ref \cite{Shmatov}. One advantage of quasimonoenergetic
ion beams is the possibility to place the source far from the
compressed core avoiding the use of a re-entrant cone. However,
the feasibility of the carbon ion fast ignition scheme (CFI)
relies on the demonstration of conversion efficiencies
comparable to those found for protons \cite{Fernandez, Badziak}.
Here, we assume a laser to quasimonoenergetic carbon ions
conversion efficiencies of about 10\% and compare the potential
of the PFI and CFI schemes for fast ignition.

The paper is organized as follows. The simulation
model is briefly summarized in Sec. II.  Energy deposition
and ignition energies of proton and carbon ions with
different deuterium-tritium (DT) density distributions
are studied in Sec. III. Ignition energies of ions and
relativistic electrons are compared in Sec. IV. Finally,
conclusions and future developments are summarized in
Sec. V.

\begin{figure}
\includegraphics[width=.47\textwidth]{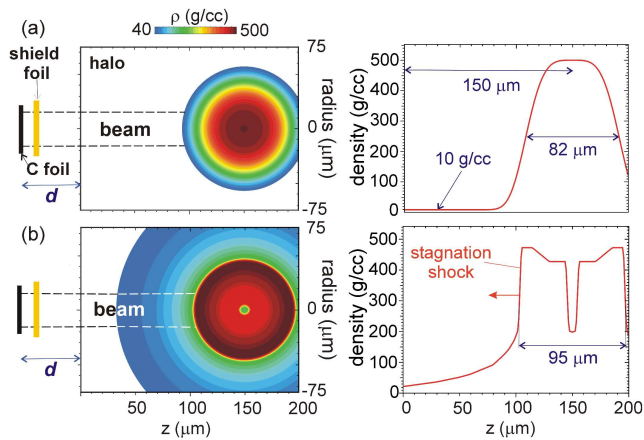}
\caption{\label{fig:1} Density maps and radial profiles
of the pre-compressed targets considered. (a) Super-Gaussian
density distribution of the spherical blob, and (b) density distribution
taken from Fig.~8 of Ref. \cite{Clark}. $d$ is the distance from the ion
source to the simulation box.}
\end{figure}

\section{Simulation model}

	We assume perfectly collimated cylindrical ion beams of
30 $\mu$m diameter impinging on a DT blob with a peak density
of 500 g/cm$^3$. Recent implosion calculations have
shown that it is possible to achieve such high densities
in capsules of 0.59 mg of DT with laser energies around
200 kJ \cite{Dunne2, Atzeni}. It is worth recalling that
densities higher than 500 g/cm$^3$ were obtained in direct-drive
implosions conducted by 10 kJ laser pulses \cite{Azechi}. We consider
the two different configurations of the compressed DT shown in Fig.~\ref{fig:1}.
The first one is a simplification of the configuration obtained from
the two-dimensional (2D) implosion calculations of cone-targets presented
elsewhere \cite{Atzeni, Honrubia1}. The second one is similar to that
obtained by Clark and Tabak \cite{Clark} from one-dimensional (1D)
implosion calculations to get an almost isochoric configuration of the compressed
fuel. As shown in Ref. \cite{Clark}, this profile corresponds to the time of maximum
density (t = 35.9 ns in Fig.~8 of \cite{Clark}) of a target with 0.9 mg of DT
compressed by a laser pulse of 480 kJ. The main differences between the two
density distributions shown in Figs.~\ref{fig:1}(a) and (b) are the higher
$\rho$R of the plasma surrounding the dense core and the steeper density ramp
of the Clark \& Tabak's target. Since the DT
plasma is stagnated at the time of peak $\rho R$, we assume that
the DT is initially at rest with an uniform temperature of 300 eV,
with exception of the central dip of the target shown in Fig.~\ref{fig:1}(b),
that has an initial temperature around 1 keV to get pressure balance
between the hot spark and the dense fuel.

Calculations have been performed with the 2-D radiation-hydrodynamics
code SARA, that includes flux-limited electron conduction, multigroup
radiation transport, ion energy deposition, DT fusion reactions and
$\alpha$-particle transport \cite{Honrubia2, Honrubia3}. We 
validated our simulation model by comparing the 
ignition energy ($\approx$ 9 kJ) of monoenergetic proton beams impinging
on an uniform DT plasma precompressed to 400 g/cm$^3$ with that obtained
by Atzeni et al.~\cite{Atzeni3} ($\approx$ 8.5 kJ).

\subsection{Ion pulse on target}

The simulation box is shown in Fig.~\ref{fig:1}.
Ions come from the left and propagate towards the blob through a
low density plasma. We assume that ions are generated instantaneously
with a Maxwellian energy distribution for protons and a Gaussian
energy distribution with a given energy spread for carbon ions.
If the energy spread is 10\% (full width at half-maximum, FWHM),
we refer to this last distribution as "quasimonoenergetic"
throughout the paper. Instantaneous emission of the beam ions is
assumed because the spread of the time of flight ($\approx$ 10 ps)
from the source to the target is much longer than the typical
ion acceleration time ($\approx$ 1 ps).

\begin{figure}
\includegraphics[width=.4\textwidth]{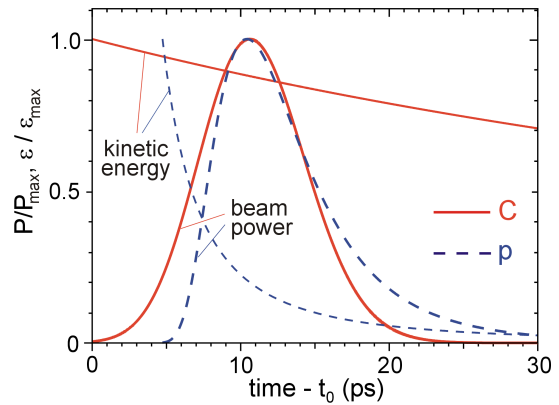}
\caption{\label{fig:2} Beam power and ion kinetic
energy at the left surface of the simulation box as
a function of time. The dashed lines correspond
to a Maxwellian proton beam with $T_p$ = 4 MeV, $d$ = 0.5 mm,
t$_0$ = 0, $\epsilon_{max}$ = 58 MeV and P$_{max}$ = 1.12 PW.
The solid lines correspond to a Gaussian carbon ion beam
with $\delta \epsilon/\epsilon_0$ = 0.1, $d$ = 1.35 cm, t$_0$ = 158 ps,
$\epsilon_{max}$ = 458 MeV and P$_{max}$ = 1.12 PW.}
\end{figure}

We used analytical formulas \cite{Temporal1, Temporal2} to compute the
kinetic energy and the beam power on target of ions accelerated
instantaneously at a distance $d$. Thus, ions with a Gaussian energy
distribution $N(\epsilon) = N_0 \sqrt{\alpha}\exp{\{-\alpha[(\epsilon-\epsilon_0)
/\Delta]^2\}}/\Delta \sqrt{\pi}$ have a power on target

\begin{equation}
 P(t) = \frac{N_0 \sqrt{\alpha}d^4m_i^2}{2\Delta \sqrt{\pi}t^5}
               \exp{\{-\alpha[d^2m_i/2\Delta t^2-\epsilon_0/\Delta]^2\}},
\end{equation}

where $N_0$ is the total number of ions ($E/\epsilon_0$),
E the beam energy, $\epsilon_0$ the mean ion kinetic energy,
$\Delta$ the energy spread (FWHM), $m_i$ the ion rest mass
and $\alpha=4\ln(2)$. Ion kinetic energy on target is given by
$\epsilon(t)=1/2m_ic^2(d/ct)^2$.
Ions with a Maxwellian energy distribution
$N(\epsilon) = 2 N_0 \sqrt{\epsilon}\exp{[-(\epsilon/kT)]}/
\sqrt{\pi}(kT)^{3/2}$ have a power on target
\begin{equation}
 P(t) = \frac{8 E}{3\sqrt{\pi}\tau}(\frac{\tau}{t})^6 \exp{\{-(\tau/t)^2}\},
\end{equation}
where E is the beam energy, $\tau$ = $d/(2kT/m_i)^{1/2}$
the pulse time scale and $T$ the ion temperature.

Beam power and ion kinetic energy of 10 kJ Maxwellian proton beam
generated at $d$ = 0.5 mm and quasimonoenergetic carbon ion
beams generated at $d$ = 1.35 cm are shown in Fig.~\ref{fig:2}.
These two distances have been chosen in order to have the same
peak power and approximately the same pulse duration on target
($\approx$ 8 ps) with both beams. It is worth noting that the
almost negligible time spread of ions with quasimonoenergetic
energy distributions allows to place the source at much higher
distances $d$ than Maxwellian ions. Because beam focalization
over distances of a few centimeters may be difficult, a number
of focusing techniques have been proposed over
the last years. Some of these techniques are: i) ballistic
transport \cite{Patel, Key}, ii) focusing by fields self-generated
in hollow microcylinders by intense sub-picosecond laser pulses
\cite{Toncian} and iii) focusing by magnetic lenses \cite{Schollmeier}.

\begin{figure}
\includegraphics[width=.4\textwidth]{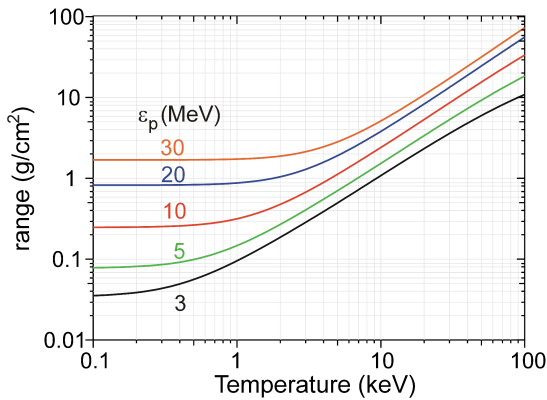}
\caption{\label{fig:3} Range of monoenergetic protons with a
kinetic energy $\epsilon_p$ versus plasma temperature in DT at
400 g/cm$^3$.}
\end{figure}

\subsection{Range lengthening}

\begin{figure}
\includegraphics[width=.4\textwidth]{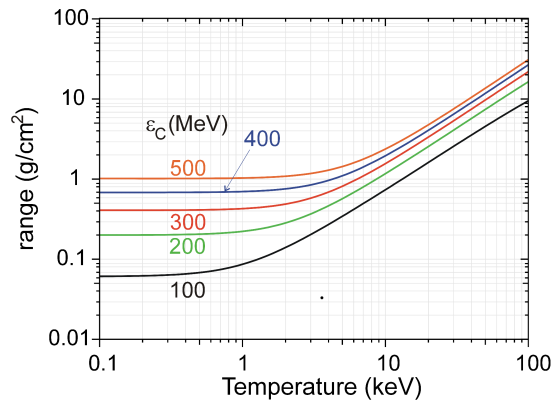}
\caption{\label{fig:4} Range of monoenergetic carbon ions with a
kinetic energy $\epsilon_C$ versus plasma temperature in DT
at 400 g/cm$^3$.}
\end{figure}

	Range of proton and carbon ions for different kinetic
energies as a function of plasma temperature are shown in Figs.~\ref{fig:3}
and \ref{fig:4}, respectively. Calculations have been performed by
pursuing the standard stopping power formalism \cite{Honrubia3}.
Ion range increases when plasma electron velocities become comparable
to the ion velocity. This effect is important for ions
with Maxwellian energy distributions placed far
from the fuel: indeed the decrease of ion kinetic energy
with time (see Fig.~\ref{fig:2}) is balanced by their
range lengthening as the DT is heated up, keeping the ion
range almost constant with time \cite{Temporal1}.
On the contrary, range lengthening may be a disadvantage
for quasimonoenergetic ions because their kinetic
energy on target has only a small variation with time
when compared with Maxwellian ions (see Fig.~\ref{fig:2}).
However, range lengthening is less pronounced for
ions with high kinetic energies per nucleon, as shown
in Figs.~\ref{fig:3} and \ref{fig:4}. For instance, the
range of 400 MeV carbon ions or 30 MeV of protons increases
only by a factor of $\approx$ 3 for plasma temperatures
from hundreds of eV to 10 keV while the range of 5 MeV protons
increases by a factor of 20 for the same temperature interval.
Nevertheless, unlike 400 MeV carbon ions, the range of
30 MeV protons at high plasma temperatures becomes much
larger than the typical value of $\approx$ 1 g/cm$^2$
required for fast ignition, increasing considerably the
ignition energy.
In general, heavier ions carry more energy per unit of mass
than protons for a given range, allowing target ignition
with much lower ion number
and beam currents.

\section{Results}

\subsection{Energy deposition}

\begin{figure}
\includegraphics[width=.45\textwidth]{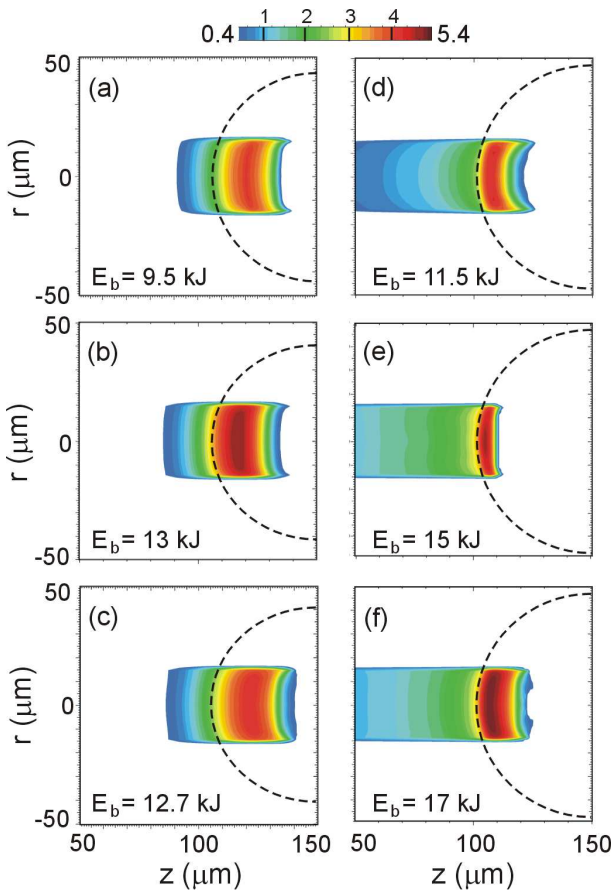}
\caption{\label{fig:5} Energy density in units of
10$^{11}$ J/cm$^3$ deposited by (a and d) a
quasimonoenergetic carbon ion beam with mean kinetic
energy of 400 MeV and $\delta \epsilon/\epsilon_0$ = 0.1
generated at a distance $d=$ 1.35 cm; (b and e) carbon
beam with Maxwellian energy distribution and
temperature $T_C$ = 100 MeV generated at $d=$ 0.5 mm;
and (c and f) proton beam with Maxwellian energy
distribution and temperature $T_p$ = 4 MeV generated at
$d=$ 0.5 mm. The left panels correspond to the target with
the supergaussian density distribution and the right panels
to the imploded target with the almost isochoric fuel
configuration. Dashed lines show the initial position
of the $\rho$ = 200 g/cm$^3$ isocontour. The total ion
beam energy E$_b$ used in each simulation is shown.}
\end{figure}

The energy deposited by proton and carbon ions with
different energy distributions in different target
configurations are compared in Fig.~\ref{fig:5}.
Maxwellian ions are generated at a distance $d$ = 0.5 mm 
from the target in order to limit their time spread.
Left panels show the energy deposition in the target 
characterized by the super-Gaussian density distribution
of Fig.~\ref{fig:1}(a). Most of the beam energy is
deposited within the dense core in a volume determined
by the ion spectrum and the range lengthening effect.
As shown in Figs. \ref{fig:5}(a-c), quasimonoenergetic
beams have a more concentrated energy deposition
than Maxwellian ions allowing for a coupling
efficiency as high as $\eta_c$ = 75\% (defined as
the fraction of the beam energy deposited at plasma
densities $\rho \ge$ 200 g/cm$^3$). Maxwellian ions
have a more distributed energy deposition and lower
coupling efficiencies, e.g. 62\% for Maxwellian carbon
ions and 65\% for Maxwellian protons. The energy
density maps shown in Figs. \ref{fig:5}(a-c) have
been obtained for beam energies $E_b$ equal to the
minimum ignition energies $E_{ig}$. These last
energies are $E_{ig}$ = 9.5 kJ for quasimonoenergetic
carbon ions, 13 kJ for Maxwellian
carbon ions with a temperature $T_C$ = 100 MeV and
12.7 kJ for Maxwellian protons with $T_p$ = 4 MeV.
Here, temperatures $T_C$ and $T_p$ are the optimal
beam temperatures for which the minimum ignition
energies are obtained. It is worthwhile noting that
despite the peak values of the energy densities
shown in Figs. \ref{fig:5}(a-c) are around
4.6$\times$10$^{11}$ J/cm$^3$, the ion beam energies
necessary to obtain them are quite different due
to the different coupling efficiencies
found for each beam. It is also remarkable that
Maxwellian protons and carbon ions have very similar
ignition energies for the optimal beam temperatures,
while the energy for quasimonoenergetic carbon
ions is about 25\% lower. In general, quasimonoenergetic
ions have a better coupling with the compressed core,
allowing fuel ignition with beam energies substantially
lower than those obtained for Maxwellian ions.

Right panels of Fig.~\ref{fig:5} show the energy deposition
for the more realistic, almost isochoric fuel configuration
shown in Fig.~\ref{fig:1}(b). In this case, the higher areal
density of the plasma surrounding the dense core reduces
the beam coupling efficiencies to $\eta_c$ = 40\%, 23\% and
39\% for the cases (d), (e) and (f) of Fig.~\ref{fig:5},
respectively. Thus, the ignition energies increase substantially,
varying from $E_{ig}$ = 11.5 kJ for quasimonoenergetic carbon
ions to 17 kJ for Maxwellian protons. The corresponding
energy densities are slightly higher in this case
than for the super-Gaussian
density distribution, ranging their peak values from
4.6$\times$10$^{11}$ J/cm$^3$ for quasimonoenergetic
carbon ions to 5.4$\times$10$^{11}$ J/cm$^3$ for Maxwellian
protons. The very low coupling efficiency found for Maxwellian
carbon ions with a beam temperature $T_C$ = 100 MeV leads
to an ignition energy higher than 20 kJ. This energy can be
reduced by raising the beam temperature to allow a deeper
penetration of carbon ions in the dense core. Optimizing
the beam temperature for the imploded fuel configuration
shown in Fig.~\ref{fig:1}(b), we find an optimal value of
$T_C$ = 200 MeV, for which $\eta_c$ = 52\% and $E_{ig}$ = 17.5 kJ,
which is again similar to that found for Maxwellian protons.

\subsection{Ignition energies for the super-Gaussian fuel configuration}

\begin{figure}
\includegraphics[width=.4\textwidth]{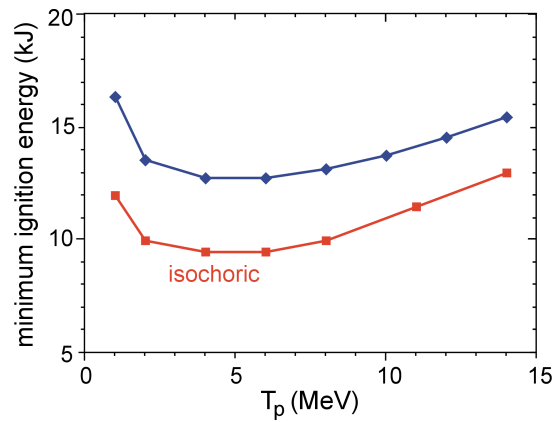}
\caption{\label{fig:6} Minimum ignition energies
of the target shown in Fig.~\ref{fig:1}(a) heated by
protons with a Maxwellian energy distribution of
temperature T$_p$. The ignition energies corresponding
to an ideal isochoric fuel are also shown. The ion source
- core distance is $d$ = 0.5 mm in all cases.}
\end{figure}

\begin{figure}
\includegraphics[width=.4\textwidth]{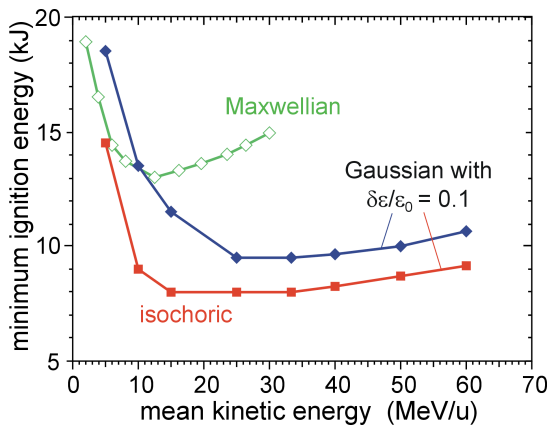}
\caption{\label{fig:7} Minimum ignition energies
of the target shown in Fig.~\ref{fig:1}(a) heated by
carbon ions with Gaussian and Maxwellian energy
distributions as a function of the mean kinetic energy
per nucleon. The ignition energies corresponding to
a Gaussian beam in an ideal isochoric fuel
are also shown. The ion source - core distances
are $d$ = 1.35 cm for the beams with Gaussian energy
distributions and $d$ = 0.5 mm for the Maxwellian one.}
\end{figure}

Ignition energies as a function of proton temperature
and kinetic energy of carbon ions are shown in Figs.
\ref{fig:6} and \ref{fig:7}, respectively. They have been
obtained as the minimum beam energy for which the
thermonuclear fusion power has an exponential or higher
growth sustained in time. Both figures show that ignition
energies increase for low and high ion kinetic energies,
reaching a minimum for intermediate values. This is
closely related with the pulses used in the simulations.
For low ion kinetic energies $\epsilon_0$, the pulse has
a relatively low power, $P_{max} \propto \epsilon_0^{1/2}$,
and long duration, $t_{pulse} \propto \epsilon_0^{-1/2}$,
as obtained from Eq. (1). In this case, the pulse parameters
depart from the optimal values and ignition energies increase.
For high ion kinetic energies, $E_{ig}$ increases again due to the
higher fuel mass heated by the ions.  We found an optimal proton
temperature $T_p$ around 4 MeV, for which $E_{ig}$ = 12.7 kJ.
This energy is higher than the $\approx$ 9 kJ found in
Ref. \cite{Temporal3} for an ideal isochoric fuel configuration
with a DT density of 500 g/cm$^3$ and the same source - core
distance $d$ = 0.5 mm. This difference accounts for the
energy deposited in the plasma surrounding the dense fuel. For
the same ideal isochoric configuration used in the reference,
our simulations give $E_{ig}$ = 9.5 kJ, in good agreement with 
\cite{Temporal3}. This result evidences the importance of the
plasma surrounding the dense fuel, that increases substantially
ignition energies. Thus, the areal density of this
plasma should be minimized by the appropriate design of the
fuel capsule or by other means, such as the two-beam scheme
discussed in Refs. \cite{Temporal2,Temporal3,Temporal4}.

Quasimonoenergetic carbon ions ($\delta \epsilon/\epsilon_0$ = 0.1)
have a better coupling with the dense fuel. Figure \ref{fig:7}
shows that ignition energies are around 9.5 kJ for the optimal
energy range from 25 to 40 MeV/u, which are about 25\% lower
than those obtained for Maxwellian protons (see Fig.~\ref{fig:6}).
Similarly to PFI, the ignition energies can be reduced
further by removing the coronal plasma surrounding the dense core.
For "ideal" isochoric targets, $E_{ig}$ is reduced to 8 kJ for
15 $-$ 32 MeV/u ions. Note that the effect of the surrounding
plasma on $E_{ig}$ is more pronounced for ion energies lower
than 25 MeV/u.

Maxwellian carbon ions have ignition energies much higher
than quasimonoenergetic ions and comparable to those
found for Maxwellian protons. For instance, Maxwellian protons
and carbon ions have ignition energies around 13 kJ for
the optimal temperatures $T_p$ = 4 MeV and $T_C$ = 100 MeV,
in agreement with the coupling efficiencies discussed in
Sec. II.A.

\begin{figure}
\includegraphics[width=.4\textwidth]{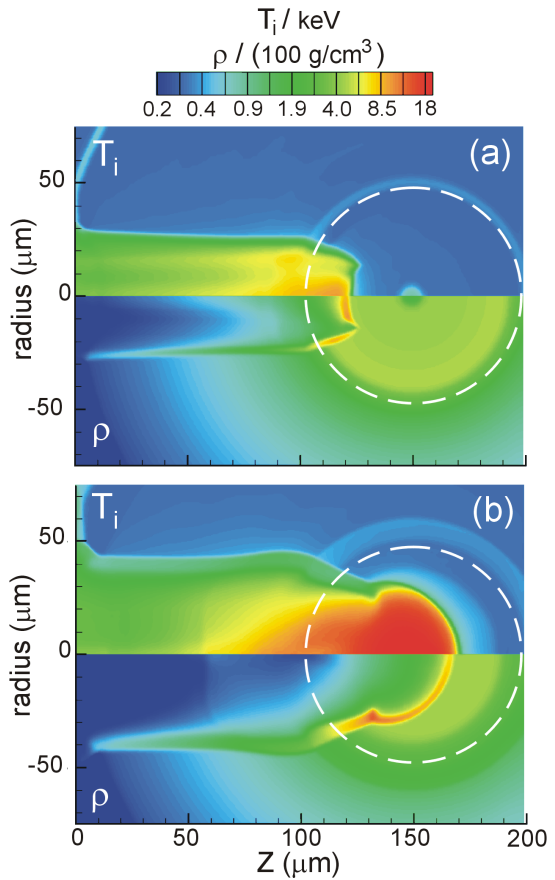}
\caption{\label{fig:8} Ion temperature ($T_i$) and density ($\rho$) 
maps of the target shown in Fig.~\ref{fig:1}(b) just after
the end of the pulse (23 ps) and (b)
when the burn wave is propagating (50 ps). The carbon ions
have a mean energy of 400 MeV with $\delta \epsilon/\epsilon_0$ = 0.1.
The total pulse energy is 12 kJ. Dashed circles show the initial
position of the $\rho$ = 200 g/cm$^3$ isocontour.}
\end{figure}

\begin{figure}
\includegraphics[width=.4\textwidth]{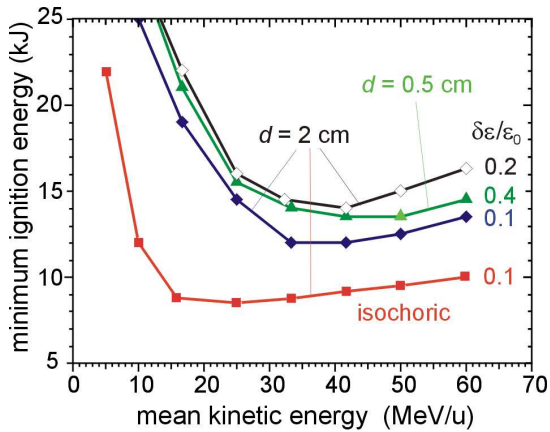}
\caption{\label{fig:9} Minimum ignition energies
of the target shown in Fig.~\ref{fig:1}(b) heated by
carbon ion beams with Gaussian energy distributions
and different energy spreads $\delta \epsilon/\epsilon_0$. The ignition
energies corresponding to an ideal isochoric fuel are
also shown. The curves labeled by 0.1 and 0.2 have been
obtained for an ion source-target distance $d=$ 2 cm, and
the curve labeled by 0.4 for $d=$ 0.5 cm.}
\end{figure}

\subsection{Ignition energies for the imploded fuel configuration}

We consider in this section quasimonoenergetic carbon ions
with different mean kinetic energies $\epsilon_0$ and energy spreads
$\delta \epsilon/\epsilon_0$ heating the imploded fuel configuration
shown in Fig.~\ref{fig:1}(b). Typical fuel density and ion temperature
maps at the end of the pulse and during the burn wave propagation
are shown in Fig.~\ref{fig:8}. Ion temperatures around 10 keV in a
hot spot of approximately 20 $\mu$m diameter can be observed in
Fig.~\ref{fig:8}(a).

	The ignition energies are shown in Fig.~\ref{fig:9}. It is remarkable
that these energies are at least 25\% higher than those obtained for the
target with the super-Gaussian density distribution (see Fig.
\ref{fig:7}). This is mainly due to the energy deposition in the plasma
surrounding the core, as can be evidenced by comparing the ignition
energies with those obtained for the ideal isochoric configuration.
Note that the effect of the surrounding plasma on $E_{ig}$ is particularly
important for relatively low ion kinetic energies. Thus, the optimal energy
range of quasimonoenergetic carbon ions is shifted from the 25 $-$ 40 MeV/u
obtained for the super-Gaussian density distribution to the 33 $-$ 50 MeV/u
shown by the curve labeled with $\delta \epsilon/\epsilon_0$ = 0.1 in
Fig.~\ref{fig:9}. 

	The effect of the ion source - core distance $d$ on the ignition
energies can be observed by comparing the curves labeled by 'isochoric' in
Figs. \ref{fig:7} and \ref{fig:9}. The lowest ignition energy changes only
slightly, from 8 kJ in Fig.~\ref{fig:7} to 8.5 kJ in Fig.~\ref{fig:9}, when
the distance $d$ increases from 1.35 cm to 2 cm, respectively. Its effect
is more pronounced, however, for ions with energies $\epsilon_0 <$ 10 MeV/u.

Ignition energies are also sensitive to the ion beam energy spread,
as can be seen by comparing the curves for $\delta \epsilon/\epsilon_0$
= 0.1 and 0.2 in Fig.~\ref{fig:9}. Spreads higher than 0.2 lead to very high
ignition energies. For instance, the ignition energies for
$\delta \epsilon/\epsilon_0$ = 0.4 and a distance $d$ = 2 cm become
higher than 20 kJ. This energy can be reduced by placing the ion
source closer to the target. Thus, the ignition energy of 50 MeV/u carbon
ions with $\delta \epsilon/\epsilon_0$ = 0.4 generated
at a distance $d$ = 0.5 cm turns out to be $E_{ig}$ = 13.5 kJ, slightly
higher than the 12.5 kJ obtained for ions with the same kinetic energy,
$\delta \epsilon/\epsilon_0$ = 0.1 and $d$ = 2 cm. Yet, a source - core
distance $d$ = 0.5 cm is still high enough to use target designs without
a re-entrant cone inserted. Therefore, we can conclude that the requirement of
quasimonoenergetic spectra can be relaxed in such a manner
that ion beams with broad energy distributions generated relatively
far from the compressed core can ignite realistic imploded
fuel configurations with moderate beam energies. If the plasma
surrounding the dense core is minimized or removed, these ignition
energies could be reduced further.

\section{Comparison with relativistic electron beams}

\begin{figure}
\includegraphics[width=.4\textwidth]{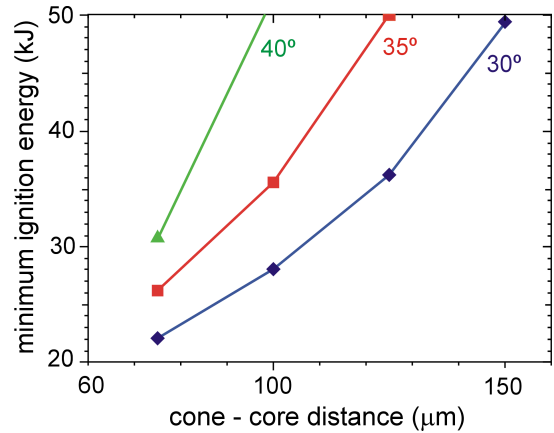}
\caption{\label{fig:10} Minimum ignition energies
of an electron beam with mean kinetic energy of
1.6 MeV impinging on the target shown in Fig.
\ref{fig:1}(a). Cone - core distance is
the distance between the left surface of the
simulation box and the center of the super-Gaussian
density distribution. The curves are labeled
with the initial divergence half-angle of the
relativistic electron beam.}
\end{figure}

Recent results on electron-driven fast ignition with a fuel
configuration similar to that depicted in Fig.~\ref{fig:1}(a)
have shown that ignition energies depends strongly on
beam divergence and distance from the cone-tip to the dense
fuel (150 $\mu$m in Fig.~\ref{fig:1}(a)) \cite{Honrubia4}.
We found that beam focusing by self-generated magnetic fields
plays an important role, reducing substantially the ignition
threshold. The ignition energies for 1.6 MeV electrons
are plotted in Fig.~\ref{fig:10}. Targets with cone - core
distances lower than 125 $\mu$m can be ignited by electron
beams with energies around 40 kJ if the beam divergence
half-angle at the cone tip is lower than 30 $-$ 35$^{\circ}$.
Assuming a laser-to-fast electron conversion efficiency of 40\%,
this energy corresponds to an energy of the multi-petawatt
laser system around 100 kJ.

Our results on ion fast ignition show that the target
pictured in Fig.~\ref{fig:1}(a) can be ignited by a Maxwellian
proton beam of 12.7 kJ with the optimal temperature $T_p$ = 4 MeV.
This energy falls to 9.5 kJ for quasimonoenergetic carbon ions
in the energy range of 25 $-$ 40 MeV/u. Assuming a laser-to-ion
conversion efficiency of 10\%, we find that fast ignition driven by
quasimonoenergetic ions requires laser energies around 100 kJ,
similar to those found for fast ignition by relativistic electrons.
Thus, the better coupling with the plasma, the lower energy requirements
and the possibility to use of targets without cones set quasimonoenergetic
ions as a promising candidate to demonstrate fast ignition
if conversion efficiencies around 10\% can be achieved experimentally
(by means of either the laser breakout afterburner
or radiation pressure acceleration schemes).

Our results also show that the ion beam energy requirements for
more realistic imploded target configurations, such as that
shown in Fig.~\ref{fig:1}(b), are around 25\% higher
(see Fig.~\ref{fig:9}). In this case, the flexibility
offered by ion-driven fast ignition can be used to
further reduce the ignition energies. The use of multiple ion
beams \cite{Temporal3,Temporal4} or ions heavier than carbon
may be an option to reduce the laser energy requirements.
On the other hand, if the ion energy spread is higher than
the 10\% assumed here as reference, our calculations show
that there is still margin to reduce the ion source - core
distance maintaining target designs without a re-entrant
cone inserted.

\section{Conclusions}
Fast ignition driven by ions presents several advantages
over fast ignition driven by relativistic electrons such
as their well know interaction with the plasma, their
focusability and the flexibility for the fuel ignition,
such as the possibility of using multiple beams. This
last possibility is particularly useful to minimize
the areal density of the plasma surrounding the dense
core and therefore to reduce significantly the ignition
energies. The use of quasimonoenergetic ions heavier than
protons improves the beam coupling efficiency and reduces
substantially the ignition energies. In addition,
quasimonoenergetic ions allow to place the ion source
far away from the fuel capsule, simplifying target design
and fuel implosion and compression, provided that ions
can be focused onto a spot of about 30 $\mu$m over distances
of a few centimetres. This conclusion is still valid for ion energy
spreads as high as 40\%. In this case, the increase of
the ignition energy can be compensated by reducing the
source - capsule distance, which is still high enough
to avoid the use of re-entrant cones.  

Prior to the application of the fast ignition by quasimonoenergetic
ions scheme, laser to ion conversion
efficiencies of the order of 10\% have to be demonstrated
experimentally. If these conversion efficiencies are achieved, our
calculations show that the laser energy requirements are similar
to those found for fast ignition driven by relativistic electrons.
In this case, the advantages of the ion fast ignition scheme
are attractive enough to envision ion fast ignition as a candidate
to demonstrate fast ignition in future facilities such as HiPER
\cite{Dunne}

Future studies will include advanced beam configurations, realistic
beam divergences and the consideration of other ions.

\begin{acknowledgments}
One of the authors (JJH) would like to thank the fruitful discussions
and the hospitality of the P-24 group of LANL. MT has been
supported by a contract of the Spanish Ministry of Education
(Ram\'on y Cajal 2007-0447).
This work was partially supported by the research grants
ENE2006-06339 and CAC-2007-013 of the Spanish Ministry of Education
and by the IFE Keep-in-touch Activities of EURATOM.
\end{acknowledgments}

\end{document}